\documentstyle[manuscript,eqsecnum,aps,tighten]{revtex}
\begin{document}
\draft
\preprint{WUHEP/97-15}

\title{Model of supersymmetric quantum field theory with broken parity symmetry}

\author{Carl M. Bender\thanks{Electronic address: cmb@howdy.wustl.edu}}
\address{Department of Physics, Washington University, St. Louis, MO 63130, USA}

\author{Kimball A. Milton\thanks{Electronic address: milton@mail.nhn.ou.edu}}
\address{Department of Physics and Astronomy, University of Oklahoma, Norman, OK
73019, USA}

\date{\today}
\maketitle

\begin{abstract}
Recently, it was observed that self-interacting scalar quantum field theories
having a non-Hermitian interaction term of the form $g(i\phi)^{2+\delta}$, where
$\delta$ is a real positive parameter, are physically acceptable in the sense
that the energy spectrum is real and bounded below. Such theories possess PT
invariance, but they are not symmetric under parity reflection or time reversal
separately. This broken parity symmetry is manifested in a nonzero value for
$\langle\phi\rangle$, even if $\delta$ is an even integer. This paper extends
this idea to a two-dimensional supersymmetric quantum field theory whose
superpotential is ${\cal S}(\phi)=-ig(i\phi)^{1+\delta}$. The resulting quantum
field theory exhibits a broken parity symmetry for all $\delta>0$. However,
supersymmetry remains unbroken, which is verified by showing that the
ground-state energy density vanishes and that the fermion-boson mass ratio is
unity.
\end{abstract}

\section{INTRODUCTION}
\label{s1}

It is extremely difficult to perform conventional perturbative calculations in
supersymmetric field theory because expansions in powers of the coupling
constant are infrared divergent. Furthermore, introducing a regulator in the
form of a momentum cutoff or a lattice spacing to control this divergence breaks
the supersymmetry invariance. An especially simple way to solve this problem is
to use the delta expansion, a perturbative expansion in powers of the degree of
the nonlinearity of the interaction term \cite{DELTA}.

The key idea of the delta expansion is to replace a self-interaction term of the
form $g\phi^4$ by $g(\phi^2)^{1+\delta}$, where $\delta$ is regarded as a small
parameter. Thus, the parameter $\delta$ measures the departure from linearity of
the field equation. A graphical procedure for expanding the Green's functions of
a quantum field theory in powers of $\delta$ is given in Ref.~\cite{DELTA}. The
advantage of such an expansion is that it has a nonzero radius of convergence
and that it yields numerically accurate nonperturbative information when
$\delta$ is set equal to 1.

The delta expansion is broadly useful for nonlinear problems. It has been
applied to many classical nonlinear ordinary differential equations of
mathematical physics and it has given superbly accurate numerical results
\cite{ODE}. The delta expansion has also been successfully applied to some
nonlinear partial differential equations of mathematical physics \cite{PDE}.

In the context of quantum field theory the delta expansion has been used to
study renormalization \cite{REN}, local gauge invariance \cite{QED}, stochastic
quantization \cite{STOCH}, finite-temperature field theory \cite{HOT}, and the
Ising limit of quantum field theory \cite{ISING}.

The delta expansion is particularly well suited for studying supersymmetric
models of quantum field theory because expansions in powers of $\delta$ are not
infrared divergent and the delta expansion respects the supersymmetry exactly.
Calculations to second order in powers of $\delta$ have been done for the
ground-state energy \cite{SUPER1} and the fermion-boson mass ratio
\cite{SUPER2}.

Recently, we have examined a new class of scalar quantum field theories that
are not symmetric under parity reflection \cite{PARITY}. The Euclidean-space
Lagrangian for this class of theories is
\begin{eqnarray}
{\cal L}={1\over2}(\partial\phi)^2+{1\over2}m^2\phi^2-g(i\phi)^{2+\delta}\qquad
(\delta>-2).
\label{e1.1}
\end{eqnarray}
The Hamiltonian for such theories is not Hermitian. However, there is strong
evidence that these theories possess energy spectra that are real and bounded 
below.\footnote{There is copious numerical evidence that such theories possess
positive spectra when $m\neq0$. Furthermore, one can understand positivity from
a theoretical point of view. Consider, for example, the weak-coupling expansion
for the $d$-dimensional Euclidean quantum field theory defined by the Lagrangian
${\cal L}={1\over2}(\partial\phi)^2+{1\over2}m^2\phi^2+gi\phi^3$. For a 
conventional $g\phi^3$ theory the weak-coupling expansion is real, and (apart
from a possible overall factor of $g$) the Green's functions are formal power
series in $g^2$. These series are not Borel summable because they do not
alternate in sign. Nonsummability reflects the fact that the spectrum of the 
underlying theory is not bounded below. However, when we replace $g$ by $ig$, 
the perturbation series remains real but now alternates in sign. Thus, the 
perturbation series is now summable and this suggests that the underlying theory
has a real positive spectrum.} The theories in Eq.~(\ref{e1.1}) are a natural
field theoretic generalization of a remarkable quantum mechanical Hamiltonian
studied by D.~Bessis and C.~Itzykson:
\cite{BESSIS}
\begin{eqnarray}
H={1\over2}p^2+ix^3.
\label{e1.2}
\end{eqnarray}
The Lagrangian in Eq.~(\ref{e1.1}) is intriguing because it is not parity
symmetric. This is manifested by a nonzero value of $\langle\phi\rangle$. It is
interesting that this broken symmetry persists even when $\delta$ is an even
integer \cite{PARITY}.

That fact that the Lagrangian in Eq.~(\ref{e1.1}) has a broken parity symmetry
suggests that if we construct a two-dimensional supersymmetric quantum field
theory by using a superpotential of the form
\begin{eqnarray}
{\cal S}(\phi)=-ig(i\phi)^{1+\delta},
\label{e1.3}
\end{eqnarray}
the resulting theory will also have a broken parity symmetry. The supersymmetric
Lagrangian resulting from the superpotential (\ref{e1.3}) is
\begin{eqnarray}
{\cal L} &=& {1\over 2}(\partial\phi)^2+{1\over 2}i\bar\psi\partial {\!\!\!/}
\psi+{1\over 2}{\cal S}'(\phi)\bar\psi\psi+{1\over 2}[{\cal S}(\phi)]^2
\nonumber\\
&=& {1\over 2}(\partial\phi)^2+{1\over 2}i\bar\psi\partial {\!\!\!/}\psi+
{1\over 2}g(1+\delta)(i\phi)^{\delta}\bar\psi\psi-{1\over 2}
g^2(i\phi)^{2+2\delta},
\label{e1.4}
\end{eqnarray}
where $\psi$ is a Majorana spinor.

The Lagrangian (\ref{e1.4}) raises an interesting question. Will the breaking of
parity symmetry induce a breaking of supersymmetry? To answer this question we
calculate both the ground-state energy $E_{\rm ground~state}$ and the
fermion-boson mass ratio $R$ as series in powers of the parameter $\delta$. We
find that through second order in $\delta$, $E_{\rm ground~state}=0$ and $R=1$,
which strongly suggests that supersymmetry remains unbroken. Based on the
experience with these calculations we believe that our results are valid to all
orders in powers of $\delta$. It is quite difficult to break supersymmetry
\cite{ZZ}.

This paper is organized very simply. In Sec.~\ref{s2} we explain our
calculational procedure. We derive a set of Feynman rules for obtaining the
Green's functions to a given order in powers of $\delta$ for the scalar field
theory Lagrangian (\ref{e1.1}). Next, in Sec.~\ref{s3} we apply the procedures
of Sec.~\ref{s2} to calculate the ground-state energy and the fermion-boson mass
ratio to first order in $\delta$ for the supersymmetric Lagrangian (\ref{e1.4}).
In Sec.~\ref{s4} we perform these calculations to second order in $\delta$.
Finally, in Sec.~\ref{s5} we examine some of the formal cancellations that occur
in our calculations to see if there could be anomalous contributions to the
ground-state energy.

\section{Delta expansion for a parity-violating scalar field theory}
\label{s2}

In this section we explain how to calculate the Green's functions for the scalar
quantum field theory described by the Lagrangian ${\cal L}$ in Eq.~(\ref{e1.1}).
Specifically, we follow the procedures described in Ref.~\cite{DELTA} and
determine a set of graphical rules for constructing the delta expansion through
second order in powers of $\delta$. These rules consist of the amplitudes for
the vertices and lines of a related Lagrangian in which the boson fields are
raised to integer powers. 

We begin by expanding the Lagrangian ${\cal L}$ in Eq.~(\ref{e1.1}) to second
order in powers of $\delta$:
\begin{eqnarray}
{\cal L}={1\over2}(\partial\phi)^2+{1\over2}m^2\phi^2+g\phi^2+\delta g\phi^2
\ln(i\phi)+{1\over2}\delta^2g\phi^2[\ln(i\phi)]^2+{\rm O}(\delta^3).
\label{e2.1}
\end{eqnarray}
Observe that at $\delta=0$ the Lagrangian contains the coupling $g$:
\begin{eqnarray}
{\cal L}_0={1\over2}(\partial\phi)^2+{1\over2}m^2\phi^2+g\phi^2.
\label{e2.2}
\end{eqnarray}
Thus, an expansion in powers of $\delta$ is clearly nonperturbative in $g$.

In general, the functional integral representation for an $n$-point correlation
function has the form (modulo the usual normalization factor)
\begin{eqnarray}
\langle 0|\phi(x_1)\phi(x_2)\phi(x_3)\ldots\phi(x_n)|0\rangle=
\int D\phi\,\phi(x_1)\phi(x_2)\phi(x_3)\ldots\phi(x_n) e^{-\int dx\,{\cal L}}.
\label{e2.3}
\end{eqnarray}
We expand the exponential in the integrand in Eq.~(\ref{e2.3}) and obtain
\begin{eqnarray}
e^{-\int dx\,{\cal L}}=e^{-\int dx\,{\cal L}_0}{\cal E}[\phi],
\label{e2.4}
\end{eqnarray}
where
\begin{eqnarray}
{\cal E}[\phi] &=& 1-\delta g\int dx\,\phi^2\ln(i\phi)-{1\over2}\delta^2 g
\int dx\,\phi^2[\ln(i\phi)]^2\nonumber\\
&&\qquad +{1\over2}\delta^2 g^2\left[\int dx\,\phi^2\ln(i\phi)\right]^2
+{\rm O}(\delta^3).
\label{e2.5}
\end{eqnarray}

Next, we use the identity
\begin{eqnarray}
\ln(i\phi)=\ln(|\phi|)+{1\over2}i\pi{\phi\over|\phi|},
\label{e2.6}
\end{eqnarray}
where $\phi/|\phi|$ represents the algebraic sign of $\phi$, to decompose the
expression for ${\cal E}$ in (\ref{e2.5}) into its real and imaginary parts:
\begin{eqnarray}
{\cal E}[\phi]=1 &-& \delta g\int dx\,\phi^2\ln(|\phi|)-{1\over2}i\pi\delta g
\int dx\,\phi|\phi|-{1\over2}\delta^2 g\int dx\,\phi^2[\ln(|\phi|)]^2\nonumber\\
&-& {1\over2}i\pi\delta^2 g\int dx\,\phi|\phi|\ln(|\phi|)+{1\over8}\pi^2\delta^2
g\int dx\,\phi^2+{1\over2}\delta^2 g^2\left[\int dx\,\phi^2\ln(|\phi|)\right]^2
\nonumber\\
&+& {1\over2}i\pi\delta^2 g^2\int dx\,\phi^2\ln(|\phi|)\int dx\,\phi|\phi|
-{1\over8}\pi^2\delta^2 g^2\left[\int dx\,\phi|\phi|\right]^2+{\rm O}(\delta^3).
\label{e2.7}
\end{eqnarray}

It is easy to construct an effective Lagrangian $\tilde{\cal L}$ having {\sl
polynomial} interaction terms that can be used to calculate the Green's
functions of (\ref{e2.7}) to first order in $\delta$:
\begin{eqnarray}
\tilde{\cal L}={\cal L}_0+{1\over2}\delta g\phi^{2\alpha+2}
+{1\over2}i\pi\delta g\alpha\phi^{2\alpha+2\gamma+1}.
\label{e2.8}
\end{eqnarray}
To use this Lagrangian we must assume that the parameters $\alpha$ and $\gamma$
are integer. We can then read off a set of Feynman amplitudes:
\begin{eqnarray}
\mbox{boson line:}&&\quad {1\over p^2+m^2+2g},\nonumber\\
\mbox{$(2\alpha+2)$ boson vertex:}&&\quad-{1\over2}\delta g(2\alpha+2)!,
\nonumber\\
\mbox{$(2\alpha+2\gamma+1)$ boson vertex:}&&\quad -{1\over2}i\pi\delta g\alpha
(2\alpha+2\gamma+1)!.
\label{e2.9}
\end{eqnarray}

Note that the vertices are of order $\delta$. Thus, if we are calculating to
first order in $\delta$ we need only include graphs having one vertex. Once a
graphical calculation has been completed we then apply the derivative operator
\begin{eqnarray}
{\cal D}={\partial\over\partial\alpha},
\label{e2.10}
\end{eqnarray}
followed by setting $\alpha=0$ and $\gamma={1\over2}$. The technique of using a
derivative operator to recover the Green's functions for the Lagrangian
(\ref{e1.1}) is like the replica trick used in calculations in statistical
mechanical models. It is a standard procedure used in all papers on the delta
expansion and is is discussed in great detail in
Refs.~\cite{DELTA,SUPER1,SUPER2}.

To perform calculations to second order in $\delta$ we seek a higher-order
effective Lagrangian $\tilde{\cal L}$ of the form
\begin{eqnarray}
\tilde{\cal L}={\cal L}_0 &+& (\delta A_1+\delta^2 A_2)g\phi^{2\alpha+2}
+(\delta B_1+\delta^2 B_2)g\phi^{2\beta+2}\nonumber\\
&+& (\delta C_1+\delta^2 C_2)g\phi^{2\alpha+2\gamma+1}
+(\delta D_1+\delta^2 D_2)g\phi^{2\beta+2\gamma+1}.
\label{e2.11}
\end{eqnarray}
To determine the coefficients $A_1$, $A_2$, $B_1$, $B_2$, $C_1$, $C_2$, $D_1$,
and $D_2$, we replace ${\cal L}$ by $\tilde{\cal L}$ on the left side of
Eq.~(\ref{e2.4}) and expand to second order in $\delta$. We then apply the
derivative operator \cite{DELTA}
\begin{eqnarray}
{\cal D}={1\over2}\left({\partial\over\partial\alpha}-{\partial\over\partial
\beta}\right)+{1\over4}\left({\partial^2\over\partial\alpha^2}+{\partial^2\over
\partial\beta^2}\right)
\label{e2.12}
\end{eqnarray}
and set $\alpha=0$, $\beta=0$, and $\gamma={1\over2}$. By comparing with the
right side of Eq.~(\ref{e2.7}) we obtain a set of seven simultaneous equations
for the coefficients:
\begin{eqnarray}
{\cal D}\Bigg( A_1\int dx\,\phi^{2\alpha+2}+B_1\int dx\,\phi^{2\beta+2}\Bigg)
\Biggm|_{\alpha=\beta=0\atop\gamma={1\over2}} &=&\int dx\,\phi^2\ln(|\phi|),
\nonumber\\
{\cal D}\Bigg( A_2\int dx\,\phi^{2\alpha+2}+B_2\int dx\,\phi^{2\beta+2}\Bigg)
\Biggm|_{\alpha=\beta=0\atop\gamma={1\over2}} &=& {1\over2}\int dx\,\phi^2
[\ln(|\phi|)]^2-{1\over8}\pi^2\int dx\,\phi^2,\nonumber\\
{\cal D}\Bigg( C_1\int dx\,\phi^{2\alpha+2\gamma+1}+D_1\int dx\,\phi^{2\beta+2
\gamma+1}\Bigg)\Biggm|_{\alpha=\beta=0\atop\gamma={1\over2}} &=&{1\over2}i\pi
\int dx\,\phi|\phi|,\nonumber\\
{\cal D}\Bigg( C_2\int dx\,\phi^{2\alpha+2\gamma+1}
+D_2\int dx\,\phi^{2\beta+2\gamma+1}\Bigg)\Biggm|_{\alpha=\beta=0\atop
\gamma={1\over2}} &=& {1\over2}i\pi\int dx\,\phi|\phi|\ln(|\phi|),\nonumber\\
{\cal D}\Bigg( A_1\int dx\,\phi^{2\alpha+2}+B_1\int dx\,\phi^{2\beta+2}\Bigg)^2
\Biggm|_{\alpha=\beta=0\atop\gamma={1\over2}}&=&\Bigg[\int dx\,\phi^2\ln(|\phi|)
\Bigg]^2,\nonumber\\
{\cal D}\Bigg( C_1\int dx\,\phi^{2\alpha+2\gamma+1}+D_1\int dx\,\phi^{2\beta+2
\gamma+1}\Bigg)^2 \Biggm|_{\alpha=\beta=0\atop\gamma={1\over2}} &=& -{1\over4}
\pi^2\Bigg(\int dx\,\phi|\phi|\Bigg)^2,\nonumber\\
&& \!\!\!\!\!\!\!\!\!\!\!\!\!\!\!\!\!\!\!\!\!\!\!\!\!\!\!\!\!\!\!\!\!\!\!\!
\!\!\!\!\!\!\!\!\!\!\!\!\!\!\!\!\!\!\!\!\!\!\!\!\!\!\!\!\!\!\!\!\!\!\!\!
\!\!\!\!\!\!\!\!\!\!\!\!\!\!\!\!\!\!\!\!\!\!\!\!\!\!\!\!\!\!\!\!\!\!\!\!
\!\!\!\!\!\!\!\!\!\!\!\!\!\!\!\!\!\!\!\!
{\cal D}\Bigg( A_1C_1\int dx\,\phi^{2\alpha+2}\int dx\,\phi^{2\alpha+2\gamma+1}
+A_1D_1\int dx\,\phi^{2\alpha+2}\int dx\,\phi^{2\beta+2\gamma+1}\nonumber\\
&& \!\!\!\!\!\!\!\!\!\!\!\!\!\!\!\!\!\!\!\!\!\!\!\!\!\!\!\!\!\!\!\!\!\!\!\!
\!\!\!\!\!\!\!\!\!\!\!\!\!\!\!\!\!\!\!\!\!\!\!\!\!\!\!\!\!\!\!\!\!\!\!\!
\!\!\!\!\!\!\!\!\!\!\!\!\!\!\!\!\!\!\!\!\!\!\!\!\!\!\!\!\!\!\!\!\!\!\!\!
\!\!\!\!\!\!\!\!\!\!\!\!\!\!\!
+B_1C_1\int dx\,\phi^{2\beta+2}\int dx\,\phi^{2\alpha+2\gamma+1}
+B_1D_1\int dx\,\phi^{2\beta+2}\int dx\,\phi^{2\beta+2\gamma+1}\Bigg)
\Biggm|_{\alpha=\beta=0\atop\gamma={1\over2}}\nonumber\\
&=& {1\over2}i\pi\int dx\,\phi^2\ln(|\phi|) \int dx\,\phi|\phi|.
\label{e2.13}
\end{eqnarray}

The solution to these equations is:
\begin{eqnarray}
A_1 &=& {1\over2},\nonumber\\
B_1 &=& -{1\over2},\nonumber\\
C_1 &=& {1\over2}i\pi\alpha,\nonumber\\
D_1 &=& -{1\over2}i\pi\beta,\nonumber\\
A_2 &=& {1\over4}-{1\over8}\pi^2\alpha^2,\nonumber\\
B_2 &=& {1\over4}-{1\over8}\pi^2\beta^2,\nonumber\\
C_2 &=& {1\over4}i\pi,\nonumber\\
D_2 &=& -{1\over4}i\pi.
\label{e2.14}
\end{eqnarray}

We thus read off the Feynman amplitudes for the effective Lagrangian $\tilde
{\cal L}$ to second order in $\delta$:
\begin{eqnarray}
\mbox{boson line:}&&\quad {1\over p^2+m^2+2g},\nonumber\\
\mbox{$(2\alpha+2)$ boson vertex:}&&\quad \left(-{1\over2}\delta-{1\over4}
\delta^2+{1\over8}\pi^2\delta^2\alpha^2\right) g(2\alpha+2)!, \nonumber\\
\mbox{$(2\beta+2)$ boson vertex:}&&\quad \left({1\over2}\delta-{1\over4}
\delta^2+{1\over8}\pi^2\delta^2\beta^2\right) g(2\beta+2)!,\nonumber\\
\mbox{$(2\alpha+2\gamma+1)$ boson vertex:}&&\quad
\left(-{1\over2}i\pi\delta\alpha-{1\over4}i\pi\delta^2\right)g(2\alpha+2\gamma
+1)!,\nonumber\\
\mbox{$(2\beta+2\gamma+1)$ boson vertex:}&&\quad \left({1\over2}i\pi\delta
\beta+{1\over4}i\pi\delta^2\right)g(2\beta+2\gamma+1)!.
\label{e2.15}
\end{eqnarray}
To second order in $\delta$ one must include all graphs containing up to two
vertices and treat the parameters $\alpha$, $\beta$, and $\gamma$ as integers.
At the end of the calculation one must then apply the derivative operator
${\cal D}$ in Eq.~(\ref{e2.12}) and set $\alpha=0$, $\beta=0$,
$\gamma={1\over2}$.

In the next two sections we generalize this approach to the case of
supersymmetric Lagrangians, identify the Feynman rules for calculating
Green's functions, and use these rules to calculate the ground-state energy,
the one-point Green's function, and the fermion-boson mass ratio.

\section{First-order Calculations for the Supersymmetric Lagrangian}
\label{s3}
In this section we show how to calculate to first order in $\delta$ the 
ground-state energy and the fermion-boson mass ratio for the two-dimensional
Euclidean field theory defined by Eq.~(\ref{e1.4}). We follow the procedure
described in Sec.~\ref{s2} and obtain a set of graphical rules for constructing
the delta expansion to a given order in powers of $\delta$. These rules are the
amplitudes for vertices and lines of a related effective Lagrangian in which the
boson fields are raised to integer powers. 

Our first objective is to obtain this related Lagrangian. We begin by expanding
Eq.~(\ref{e1.4}) to first order in $\delta$:
\begin{eqnarray}
{\cal L} &=& {1\over2}(\partial\phi)^2+{1\over 2}g^2\phi^2
+{1\over 2}i\bar\psi\partial {\!\!\!/}\psi+{1\over 2}g\bar\psi\psi\nonumber\\
&&\qquad +\delta g^2\phi^2\ln(i\phi)+{1\over2}
\delta g\bar\psi\psi+{1\over2}\delta g\bar\psi\psi\ln(i\phi)+{\rm O}(\delta^2).
\label{e3.1}
\end{eqnarray}
We can display the real and imaginary parts of this Lagrangian explicitly by
using the identity (\ref{e2.6}).

Unfortunately, the Lagrangian (\ref{e3.1}) is nonpolynomial so we cannot read
off a conventional set of Feynman rules. However, consider the following
effective Lagrangian whose interaction terms are polynomial in form:
\begin{eqnarray}
\tilde{\cal L} &=& {1\over 2}(\partial\phi)^2+{1\over 2}g^2\phi^2+{1\over2}i\bar
\psi\partial{\!\!\!/}\psi+{1\over 2}g\bar\psi\psi\nonumber\\
&&\quad +{1\over2}\delta g^2\phi^{2+2\alpha}+{1\over4}\delta(2\alpha+1)g\bar\psi
\psi\phi^{2\alpha}+{1\over2}i\pi\delta g^2\alpha\phi^{2\alpha+2\gamma+1}
+{1\over4}i\pi\delta g\alpha\bar\psi\psi\phi^{2\alpha+2\gamma-1},
\label{e3.2}
\end{eqnarray}
where $\alpha$ and $\gamma$ are to be regarded temporarily as positive integers.
Note that we recover the Lagrangian ${\cal L}$ in Eq.~(\ref{e3.1}) if we apply
the derivative operator ${\cal D}$ in Eq.~(\ref{e2.10}) to $\tilde{\cal L}$ and
set $\alpha=0$ and $\gamma={1\over2}$. 

To calculate the Green's functions for ${\cal L}$ we use the same device. That
is, we find the Green's functions for $\tilde{\cal L}$. We then apply ${\cal D}$
to these Green's functions and set $\alpha=0$ and $\gamma={1\over2}$. To
calculate the Green's functions for $\tilde{\cal L}$ we read off the following
Feynman rules from Eq.~(\ref{e3.2}):
\begin{eqnarray}
\mbox{boson line:}&&\quad \tilde\Delta(p)={1\over p^2+g^2},\nonumber\\
\mbox{fermion line:}&&\quad \tilde S(p)={1\over g-p\!\!/},\nonumber\\
\mbox{$(2\alpha+2)$ boson vertex:}&&\quad-{1\over2}\delta g^2(2\alpha+2)!,
\nonumber\\
\mbox{$(2\alpha)$ boson and 2 fermion vertex:}&&\quad
-{1\over2}\delta g(2\alpha+1)!,\nonumber\\
\mbox{$(2\alpha+2\gamma+1)$ boson vertex:}&&\quad -{1\over2}i\pi\delta g^2\alpha
(2\alpha+2\gamma+1)!,\nonumber\\
\mbox{$(2\alpha+2\gamma-1)$ boson and 2 fermion vertex:}&&\quad-{1\over2}i\pi
\delta g\alpha(2\alpha+2\gamma-1)!.
\label{e3.3}
\end{eqnarray}
Note that these rules are {\em nonperturbative} in the coupling constant $g$;
the parameter $g$ appears nontrivially in all the vertex and line amplitudes.

In addition to these Feynman rules one must remember to associate a factor of 
$-1$ with every fermion loop, and that in two-dimensional space gamma matrices
are two dimensional so that the trace of the unit matrix introduces a factor of
two. Moreover, one must be careful to associate with each graph the appropriate
symmetry number. The fermions for this quantum field theory are Majorana
fermions, which are nondirectional. Thus, for example, the symmetry number of a
fermion self-loop consisting of a single fermion propagator is ${1\over2}$.

Using these graphical rules, we calculate below the zero-point, one-point, and
two-point Green's functions to order $\delta$. Because we are calculating to
first order in $\delta$ and all vertices in Eq.~(\ref{e3.3}) are of order
$\delta$, we need only consider single-vertex graphs. We will show that to order
$\delta$ the ground-state energy vanishes and that the renormalized fermion and
boson masses are equal. This indicates that the theory is supersymmetric to this
order. We also see that the one-point Green's function is nonvanishing, which
shows that parity is broken to order $\delta$. These results are obtained using
a formal manipulation of divergent quantities in the form of logarithmically
divergent integrals. In Sec.~\ref{s5} we re-examine these formal manipulations
more carefully to see whether an anomalous structure could be present.

\subsection{Calculation of the ground-state energy}
\label{ss3a}
The ground-state energy of a quantum field theory is the negative sum of the
connected graphs having no external lines. Because we are considering graphs
with no external lines, only the vertices in Eq.~(\ref{e3.3}) having an even
number of boson lines contribute to the ground-state energy. Our results can be
expressed in terms of a single dimensionless divergent integral $\Lambda$, which
represents the amplitude of a boson self-loop (a loop formed from one boson
propagator):
\begin{eqnarray}
\Lambda=\Delta(0)={1\over(2\pi)^2}\int d^2p\, {1\over p^2+g^2},
\label{e3.4}
\end{eqnarray}
where $\Delta(x)$ is the boson propagator in coordinate space [the Fourier
transform of $\tilde\Delta(p)$]. The amplitude for a fermion self-loop can also
be expressed in terms of $\Lambda$:
\begin{eqnarray}
{1\over(2\pi)^2}\int d^2p\,{\rm Tr}\left({1\over g-p\!\!/}\right)={1\over(2
\pi)^2}\int d^2p\,{\rm Tr}\left({g+p\!\!/\over p^2+g^2}\right)=2g\Lambda.
\label{e3.5}
\end{eqnarray}

Two graphs of order $\delta$ contribute to the ground-state energy density.
These graphs are shown in Fig.~\ref{fig1}. The amplitudes for these graphs are
the product of the vertex and line amplitudes multiplied by the appropriate
symmetry numbers. The first graph is a pure boson graph consisting of one vertex
with $\alpha+1$ boson self-loops attached. The amplitude for the vertex is
$-{1\over2}\delta g^2(2\alpha+2)!$, the symmetry number for this graph is
$2^{-\alpha-1}/(\alpha+1)!$, the Feynman integral for the graph is $\Lambda
^{\alpha+1}$, and we include a factor of $-1$ because we are calculating the
ground-state energy density. The second graph is a mixed boson-fermion graph
consisting of one vertex with $\alpha$ boson self-loops and one fermion
self-loop attached. The amplitude for the vertex is $-{1\over2}\delta g(2\alpha
+1)!$, the symmetry number for the graph is $2^{-\alpha-1}/\alpha!$, the Feynman
integral for the graph is $2g\Lambda^{\alpha+1}$, and there are two factors of
$-1$, the first because there is a fermion loop and the second because we are
calculating the ground-state energy density. In summary, our results are
\begin{eqnarray}
\mbox{Fig.~1a:}&&\quad\delta g^2{(2\alpha+1)!\Lambda^{\alpha+1}\over2^{\alpha
+1}\alpha!},\nonumber\\
\mbox{Fig.~1b:}&&\quad-\delta g^2{(2\alpha+1)!\Lambda^{\alpha+1}\over2^{\alpha
+1}\alpha!}.
\label{e3.6}
\end{eqnarray}

Observe that the two amplitudes above are equal in magnitude but opposite in
sign. Thus, the final result for the ground-state energy to order $\delta$
appears to be $0$. This explicit cancellation holds for {\sl all} values of
$\alpha$; although we expected to differentiate with respect to $\alpha$ and set
$\alpha=0$, there seems to be no need to do so in this case and we obtain
directly
\begin{eqnarray}
E_{\rm ground~state}=0+{\rm O}(\delta^2).
\label{e3.7}
\end{eqnarray}
Of course, the cancellation that leads to this result comes from subtracting two
logarithmically divergent integrals. One may ask whether this cancellation
persists if these integrals are properly regulated. We examine this question in
Sec.~\ref{s5}.

\subsection{Calculation of $\langle\phi\rangle$}
\label{ss3b}
In contrast, the one-point Green's function arises only from vertices in
Eq.~(\ref{e3.3}) having an odd number of boson lines. The two contributing
graphs are shown in Fig.~\ref{fig2}. The amplitudes are
\begin{eqnarray}
\mbox{Fig.~2a:}&&\quad-i\pi\delta\alpha{(2\alpha+2\gamma+1)!\,
\Lambda^{\alpha+\gamma}\over 2^{\alpha+\gamma+1}(\alpha+\gamma)!},\nonumber\\
\mbox{Fig.~2b:}&&\quad i\pi\delta\alpha{(2\alpha+2\gamma-1)!\,\Lambda^{\alpha+
\gamma}\over2^{\alpha+\gamma} (\alpha+\gamma-1)!}.
\label{e3.8}
\end{eqnarray}
Adding these amplitudes, differentiating with respect to $\alpha$, and setting
$\alpha=0$ and $\gamma={1\over2}$, we obtain a nonzero result for the vacuum
expectation value of the scalar field:
\begin{eqnarray}
\langle\phi\rangle=G_1=-i\delta\sqrt{\pi\Lambda/2}+{\rm O}(\delta^2).
\label{e3.9}
\end{eqnarray}
We conclude that this supersymmetric theory has a broken parity symmetry.

On the basis of calculations performed in Ref.~\cite{PARITY}, we believe that
$\langle\phi\rangle$ remains nonzero even when $\delta$ is a positive integer.
One might think that the Lagrangian (\ref{e1.4}) is surely parity symmetric when
$\delta$ is an even integer and one might worry the theory does not exist when
$\delta$ is an odd integer [because the term $-(i\phi)^{2\delta+2}$ appears to
be unbounded below]. However, neither of these concerns is realized. The reason
is that as $\delta$ increases from 0 (free field theory) the entire theory must
be analytically continued as a function of $\delta$. The boundary conditions on
the functional integral $Z=\int D\phi\,\exp(-\int dx{\cal L})$ representing the
vacuum persistence function rotate into the complex-$\phi$ plane and yield a
broken-parity theory that exists for all $\delta$. This analytic continuation of
boundary conditions is discussed in great detail in Ref.~\cite{PARITY}. These
arguments are based on analysis given in Ref.~\cite{ROT}.

\subsection{Calculation of the fermion-boson mass ratio}
\label{ss3c}
Next, we calculate the mass renormalization of the fermion and the boson. These
mass shifts are obtained by evaluating the negative amputated
one-particle-irreducible graphs representing to the two-point Green's functions.
One graph contributes to the mass shift of the fermion to order $\delta$ and two
graphs contribute to the mass shift of the boson to order $\delta$. These graphs
are shown in Fig.~\ref{fig3}. The symmetry number for each graph is indicated in
the figure. The negative amplitude for the fermion graph is
\begin{eqnarray}
\mbox{Fig.~3a:}\quad\delta g{(2\alpha+1)!\,\Lambda^\alpha\over2^{\alpha+1}
\alpha!}.
\label{e3.10}
\end{eqnarray}
The negative amplitudes for the boson graphs are
\begin{eqnarray}
\mbox{Fig.~3b:}&&\quad\delta g^2{(2\alpha+2)!\,\Lambda^\alpha\over2^{\alpha+1}
\alpha!},\nonumber\\
\mbox{Fig.~3c:}&&\quad-\delta g^2{(2\alpha+1)!\,\Lambda^\alpha\over2^\alpha
(\alpha-1)!}.
\label{e3.11}
\end{eqnarray}

Based on our calculational procedure, the next step would be to differentiate
each of these amplitudes with respect to $\alpha$ and set $\alpha=0$. However,
this is in fact not even necessary. For arbitrary $\alpha$ the fermion mass is
\begin{eqnarray}
M_{\rm fermion}=g+\delta g{(2\alpha+1)!\,\Lambda^\alpha\over2^{\alpha+1}
\alpha!}+{\rm O}(\delta^2)
\label{e3.12}
\end{eqnarray}
and the boson mass squared is
\begin{eqnarray}
M_{\rm boson}^2=g^2+\delta g{(2\alpha+1)!\,\Lambda^\alpha\over2^\alpha \alpha!}
+{\rm O}(\delta^2).
\label{e3.13}
\end{eqnarray}
Thus, to order $\delta$ the fermion-boson mass ratio $R$ is unity:
\begin{eqnarray}
R={M_{\rm fermion}\over M_{\rm boson}}=1+{\rm O}(\delta^2).
\label{e3.14}
\end{eqnarray}

We conclude from these calculations that while parity symmetry is clearly
broken, the supersymmetry remains to order $\delta$. In the next section we
pursue these calculations to second order in $\delta$.

\section{Second-order Calculations for the Supersymmetric Lagrangian}
\label{s4}

In this section we extend the calculations of the previous section to second
order in $\delta$. We begin by expanding the Lagrangian (\ref{e1.4}) to order
$\delta^2$:
\begin{eqnarray}
{\cal L} &=& {1\over2}(\partial\phi)^2+{1\over 2}g^2\phi^2+{1\over 2}i\bar\psi
\partial{\!\!\!/}\psi+{1\over 2}g\bar\psi\psi
+{1\over2}\delta g\bar\psi\psi\nonumber\\
&&\qquad+\delta g^2\phi^2\ln(i\phi)+{1\over2}\delta g\bar\psi\psi\ln(i\phi)
+\delta^2 g^2\phi^2[\ln(i\phi)]^2\nonumber\\
&&\qquad+{1\over4}\delta^2 g\bar\psi\psi[\ln(i\phi)]^2+{1\over2}\delta^2
g\bar\psi\psi\ln(i\phi)+{\rm O}(\delta^3).
\label{e4.1}
\end{eqnarray}
Next we substitute the identity (\ref{e2.6}) and identify a polynomial
Lagrangian that in combination with the derivative operator (\ref{e2.12}) gives
the Green's functions of the theory defined by the Lagrangian (\ref{e4.1}) to
order $\delta^2$:
\begin{eqnarray}
\tilde{\cal L} &=& {1\over 2}(\partial\phi)^2+{1\over 2}g^2\phi^2
+{1\over 2}i\bar\psi\partial{\!\!\!/}\psi+{1\over 2}g\bar\psi\psi\nonumber\\
&&~ +{1\over4}(2\delta+2\delta^2-\pi^2\delta^2\alpha^2)g^2\phi^{2+2\alpha}
-{1\over4}(2\delta-2\delta^2+\pi^2\delta^2\beta^2)g^2\phi^{2+2\beta}\nonumber\\
&&~ +{1\over16}(4\delta+8\delta\alpha+2\delta^2+4\delta^2\alpha-\pi^2\delta^2
\alpha^2)g\bar\psi\psi\phi^{2\alpha}\nonumber\\
&&~ -{1\over16}(4\delta+8\delta\beta-2\delta^2-4
\delta^2\beta+\pi^2\delta^2\beta^2)g\bar\psi\psi\phi^{2\beta}\nonumber\\
&&~+{1\over2}i\pi(\delta\alpha+\delta^2)g^2\phi^{2\alpha+2\gamma+1}
-{1\over2}i\pi(\delta\beta+\delta^2)g^2\phi^{2\beta+2\gamma+1}\nonumber\\
&&~+{1\over8}i\pi(2\delta\alpha+\delta^2\alpha+2\delta^2\alpha^2)
g\bar\psi\psi\phi^{2\alpha+2\gamma-1}
-{1\over8}i\pi(2\delta\beta-\delta^2\beta-2\delta^2\beta^2)
g\bar\psi\psi\phi^{2\beta+2\gamma-1}.
\label{e4.2}
\end{eqnarray}
To find this effective Lagrangian requires considerable algebra; one must solve
a system of eighteen simultaneous equations similar in form to those in
Eq.~(\ref{e2.13}).

In this effective Lagrangian we treat the parameters $\alpha$, $\beta$, and
$\gamma$ as positive integers so that we can derive a set of Feynman rules for
calculating the Green's functions. After these Green's functions have been
calculated to second order in $\delta$ we apply the derivative operator
${\cal D}$ in Eq.~(\ref{e2.12}) and set $\alpha=0$, $\beta=0$, and
$\gamma={1\over2}$.

The Feynman rules for the Lagrangian (\ref{e4.2}) are the generalization of the
rules in Eq.~(\ref{e3.3}) to second-order in $\delta$:
\begin{eqnarray}
\mbox{boson line:}&&\quad{1\over p^2+g^2},\nonumber\\
\mbox{fermion line:}&&\quad {1\over g-p\!\!/},
\nonumber\\
\mbox{$(2\alpha+2)$ boson vertex:}&&~v_1={1\over4}(-2\delta-2\delta^2
+\pi^2\delta^2\alpha^2)g^2(2\alpha+2)!,\nonumber\\
\mbox{$(2\beta+2)$ boson vertex:}&&~v_2={1\over4}(2\delta-2\delta^2
+\pi^2\delta^2\beta^2)g^2(2\beta+2)!,\nonumber\\
\mbox{$(2\alpha)$ boson and 2 fermion vertex:}&&~v_3=
{1\over8}[-(4+8\alpha)\delta-(2+4\alpha-\pi^2\alpha^2)\delta^2]g(2\alpha)!,
\nonumber\\
\mbox{$(2\beta)$ boson and 2 fermion vertex:}&&~v_4=
{1\over8}[(4+8\beta)\delta-(2+4\beta-\pi^2\beta^2)\delta^2]g(2\beta)!,
\nonumber\\
\mbox{$(2\alpha+2\gamma+1)$ boson vertex:}&&~v_5={i\pi\over2}
(-\delta\alpha-\delta^2)g^2(2\alpha+2\gamma+1)!,\nonumber\\
\mbox{$(2\beta+2\gamma+1)$ boson vertex:}&&~v_6={i\pi\over2}
(\delta\beta+\delta^2)g^2(2\beta+2\gamma+1)!,\nonumber\\
\mbox{$(2\alpha+2\gamma-1)$ boson and 2 fermion vertex:}&&~v_7={i\pi\alpha
\over4}[-2\delta-(1+2\alpha)\delta^2]g(2\alpha+2\gamma-1)!,\nonumber\\
\mbox{$(2\beta+2\gamma-1)$ boson and 2 fermion vertex:}&&~v_8=
{i\pi\beta\over4}[2\delta-(1+2\beta)\delta^2]g(2\beta+2\gamma-1)!.
\label{e4.3}
\end{eqnarray}

\subsection{Calculation of the ground-state energy to second order in $\delta$}
\label{ss4a}
There are thirty graphs that contribute to the ground-state energy density.
These graphs are organized into seven classes for which the amplitudes combine
in a natural way. First, we examine the four single-vertex graphs that are
constructed from vertices $v_1$, $v_2$, $v_3$, and $v_4$. These graphs are shown
in Fig.~4(a). The sum of the amplitudes for these graphs is
\begin{eqnarray}
A_1=-{v_1 \Lambda^{\alpha+1} \over 2^{\alpha+1}(\alpha+1)!}
-{v_2 \Lambda^{\beta+1} \over 2^{\beta+1}(\beta+1)!}
+{v_3g \Lambda^{\alpha+1} \over 2^{\alpha}\alpha!}
+{v_4g \Lambda^{\beta+1} \over 2^{\beta}\beta!}.
\label{e4.4}
\end{eqnarray}
When these amplitudes are combined we find that the contribution of order
$\delta$ is identically $0$. Only terms of order $\delta^2$ survive.
Simplifying the expression in Eq.~(\ref{e4.4}) gives
\begin{eqnarray}
A_1={\delta^2g^2\Lambda^{\alpha+1}(2\alpha)!\over 2^{\alpha+3}\alpha!}
[4\alpha+2-\pi^2\alpha^2(4\alpha+1)]+(\alpha\to\beta).
\label{e4.5}
\end{eqnarray}

Next, we consider the two-vertex graphs constructed from $v_1$ and $v_3$, which
are shown in Fig.~4(b). Let $2l$ be the number of boson lines connecting these
two vertices. Then, for each value of $l$ the sum of the amplitudes for these
graphs is
\begin{eqnarray}
A_2(l)=&-&{v_1^2\Lambda^{2\alpha+2-2l}\int dx\,\Delta^{2l}(x)
\over[(\alpha+1-l)!]^2 2^{2\alpha+3-2l}(2l)!}
+{v_1v_3g\Lambda^{2\alpha+2-2l}\int dx\,\Delta^{2l}(x)
\over(\alpha+1-l)!(\alpha-l)! 2^{2\alpha+1-2l}(2l)!}\nonumber\\
&-&{v_3^2g^2\Lambda^{2\alpha+2-2l}\int dx\,\Delta^{2l}(x)
\over[(\alpha-l)!]^2 2^{2\alpha+1-2l}(2l)!}
+{v_3^2\Lambda^{2\alpha-2l}\int dx\,\Delta^{2l}(x)\mbox{Tr}\,\bar S(x)S(x)
\over[(\alpha-l)!]^2 2^{2\alpha+2-2l}(2l)!}.
\label{e4.6}
\end{eqnarray}
Note that we must sum over all possible values of $l$. In all of the above terms
except the last, $l$ ranges from $1$ to $\infty$, but in the second $v_3^2$ term
$l$ ranges from $0$ to $\infty$. To simplify the Feynman integral in this
term representing the boson- and fermion-exchange graph, we use the following
identity stated in \cite{SUPER1}:
\begin{eqnarray}
\int dx\,\Delta^{l}(x)\mbox{Tr}\,\bar S(x)S(x)={2g^2(l+2)\over l+1}\int dx\,
[\Delta(x)]^{l+2}-{2\over l+1}\Lambda^{l+1},\quad l\geq 0,
\label{e4.7}
\end{eqnarray}
where $S(x)$ and $\Delta(x)$ are the fermion and boson propagators in coordinate
space:
\begin{eqnarray}
S(x)&=&(g-i\partial{\!\!\!/})\Delta(x),\nonumber\\
\Delta(x)&=&{1\over2\pi}K_0(g|x|).
\label{e4.prop}
\end{eqnarray}
When we simplify the expression $A_2(l)$ there is a remarkable cancellation that
occurs; all integrals over $\Delta^{2l}(x)$ cancel independently of the value of
$\alpha$ and we obtain the following simple sum:
\begin{eqnarray}
A_2=\sum_l A_2(l)=-g^2\delta^2\Lambda^{2\alpha+1}[(2\alpha+1)!]^2 2^{-2\alpha-5}
\sum_{l=1}^{\infty} {2^{2l}\over (2l-1)![(\alpha+1-l)!]^2}.
\label{e4.8}
\end{eqnarray}

Third, we consider the two-vertex graphs constructed from $v_2$ and $v_4$,
which are shown in Fig.~4(c). The sum of the amplitudes for these graphs is
identical to $A_2$ in Eq.~(\ref{e4.8}) with the replacement $\alpha\to\beta$:
\begin{eqnarray}
A_3=A_2\Bigm|_{\alpha\to\beta}.
\label{e4.9}
\end{eqnarray}

Fourth, we examine the two-vertex graphs constructed from one of $v_1$ and $v_3$
and one of $v_2$ and $v_4$. These graphs are shown in Fig.~4(d). For each
value of $l$ the sum of the amplitudes for these graphs is
\begin{eqnarray}
A_4(l)=&-&{v_1v_2\Lambda^{\alpha+\beta+2-2l}\int dx\,\Delta^{2l}(x)
\over (\alpha+1-l)!(\beta+1-l)!2^{\alpha+\beta+2-2l}(2l)!}
+{v_1v_4g\Lambda^{\alpha+\beta+2-2l}\int dx\,\Delta^{2l}(x)
\over(\alpha+1-l)!(\beta-l)! 2^{\alpha+\beta+1-2l}(2l)!}\nonumber\\
&+&{v_2v_3g\Lambda^{\alpha+\beta+2-2l}\int dx\,\Delta^{2l}(x)
\over(\alpha-l)!(\beta+1-l)! 2^{\alpha+\beta+1-2l}(2l)!}
-{v_3v_4g^2\Lambda^{\alpha+\beta+2-2l}\int dx\,\Delta^{2l}(x)
\over (\alpha-l)! (\beta-l)! 2^{\alpha+\beta-2l}(2l)!}\nonumber\\
&+&{v_3v_4\Lambda^{\alpha+\beta-2l}\int dx\,\Delta^{2l}(x)\mbox{Tr}\,\bar S(x)
S(x)\over (\alpha-l)! (\beta-l)! 2^{\alpha+\beta+1-2l}(2l)!}.
\label{e4.10}
\end{eqnarray}
Again, we must sum over all possible values of $l$; in all of the above terms
except the last, $l$ ranges from $1$ to $\infty$, but in the $v_3v_4$ term
representing boson- and fermion-exchange, $l$ ranges from $0$ to $\infty$. Also,
we again use the integral identity in Eq.~(\ref{e4.7}). As before, when we
simplify the expression $A_4(l)$ there is a remarkable cancellation of the
$\Delta^{2l}(x)$ integrals for all values of $\alpha$ and $\beta$. We obtain the
following simple sum:
\begin{eqnarray}
A_4=\sum_l A_4(l)=g^2\delta^2\Lambda^{\alpha+\beta+1}(2\alpha+1)!(2\beta+1)!
\sum_{l=1}^{\infty}{2^{2l-\alpha-\beta-4}\over(2l-1)!(\alpha+1-l)!(\beta+1-l)!}.
\label{e4.11}
\end{eqnarray}

Next, we consider the two-vertex graphs constructed from $v_5$ and $v_7$, which
are shown in Fig.~4(e). For each value of $l$ the sum of the amplitudes for
these graphs is
\begin{eqnarray}
A_5(l)=&-&{v_5^2\Lambda^{2\alpha+2\gamma-2l}\int dx\,\Delta^{2l+1}(x)
\over[(\alpha+\gamma-l)!]^2 2^{2\alpha+2\gamma+1-2l}(2l+1)!}\nonumber\\
&+&{v_5v_7g\Lambda^{2\alpha+2\gamma-2l}\int dx\,\Delta^{2l+1}(x)\over(\alpha+
\gamma-l)!(\alpha+\gamma-1-l)! 2^{2\alpha+2\gamma-1-2l}(2l+1)!}\nonumber\\
&-&{v_7^2g^2\Lambda^{2\alpha+2\gamma-2l}\int dx\,\Delta^{2l+1}(x)
\over[(\alpha+\gamma-1-l)!]^2 2^{2\alpha+2\gamma-1-2l}(2l+1)!}\nonumber\\
&+&{v_7^2\Lambda^{2\alpha+2\gamma-2-2l}\int dx\,\Delta^{2l+1}(x)\mbox{Tr}\,
\bar S(x)S(x)\over[(\alpha+\gamma-1-l)!]^2 2^{2\alpha+2\gamma-2l}(2l+1)!}.
\label{e4.12}
\end{eqnarray}
Again, we must sum over all possible values of $l$; in all of the above terms
$l$ ranges from $0$ to $\infty$. Also, in the boson- and fermion-exchange graph
in the last term, we again use the integral identity in Eq.~(\ref{e4.7}).
When we simplify the expression $A_5(l)$ there is no longer any cancellation
for arbitrary values of $\alpha$. Thus, the final result still contains an
integral over $\Delta^{2l+1}(x)$. We obtain the following sum:
\begin{eqnarray}
A_5=\sum_l A_5(l)&=& \pi^2 g^2\delta^2\alpha^2 \Lambda^{2\alpha+2\gamma}
[(2\alpha+2\gamma-1)!]^2 2^{-2\alpha-2\gamma-3}\Biggm\{
\sum_{l=0}^{\infty} {2^{2l}\over (2l)![(\alpha+\gamma-l)!]^2}\nonumber\\
&&+\sum_{l=0}^{\infty} {g^2 2^{2l}[(2\alpha+2\gamma)^2-1][(2\alpha+2\gamma)^2
+4l+1]\int dx\,\Delta^{2l+1}(x)\over (2l+1)![(\alpha+\gamma-l)!]^2\Lambda^{2l}}
\Biggm\}.
\label{e4.13}
\end{eqnarray}

Sixth, we consider the two-vertex graphs constructed from $v_6$ and $v_8$,
which are shown in Fig.~4(f). The sum of the amplitudes for these graphs is
identical to $A_5$ in Eq.~(\ref{e4.13}) with the replacement $\alpha\to\beta$:
\begin{eqnarray}
A_6=A_5\Bigm|_{\alpha\to\beta}.
\label{e4.14}
\end{eqnarray}

The seventh and last class of graphs is constructed from one of $v_5$ and $v_7$
and one of $v_6$ and $v_8$. These graphs are shown in Fig.~4(g). It is easy to 
see that the amplitudes for all such graphs are proportional to $\alpha\beta$.
Thus, anticipating that at the end of the calculation we will set both $\alpha
=0$ and $\beta=0$, we need not calculate the amplitude $A_7$ because it will not
contribute to the ground-state energy density. (Note that before we set $\alpha$
and $\beta$ to $0$, we must apply the differential operator ${\cal D}$ in
Eq.~(\ref{e2.12}). This operator does not have a mixed derivative term and thus
it cannot eliminate both factors of $\alpha$ and $\beta$.)

The final part of this calculation consists of applying the operator ${\cal D}$
in Eq.~(\ref{e2.12}) to $A_1+A_2+A_3+A_4+A_5+A_6+A_7$ and setting $\alpha=0$,
$\beta=0$, and $\gamma={1\over2}$. After a rather lengthy calculation, we obtain
the result
\begin{eqnarray}
&&E_{\rm ground~state}
={\cal D}\sum_{j=1}^7 A_j\Biggm|_{\alpha=\beta=0,~\gamma=1/2}\nonumber\\
&=& {1\over16}\delta^2 g^2\Lambda\left[ 2\psi'\left({3\over2}\right)-2\pi^2
+\sum_{l=0}^{\infty}{\sqrt{\pi}\,\Gamma\left(l-{1\over2}\right)\over\left(l-{1
\over2}\right)\Gamma(l+1)}-\sum_{l=1}^{\infty}{\sqrt{\pi}\,\Gamma(l)\over l\,
\Gamma\left(l+{3\over2}\right)}\right]+{\rm O}(\delta^3).
\label{e4.15}
\end{eqnarray}
The sums in this expression may be evaluated easily and we obtain the result
\begin{eqnarray}
E_{\rm ground~state}=0+{\rm O}(\delta^3).
\label{e4.16}
\end{eqnarray}

\subsection{Calculation of $\langle\phi\rangle$ to second order in $\delta$}
\label{ss4b}
The graphs that contribute to the vacuum expectation value of the scalar field
arise either from one-vertex graphs constructed from the odd vertices, $v_5$
through $v_8$, or from two vertex graphs with one odd vertex and one even 
vertex. Some of these graphs are shown in Fig.~\ref{fig5}. Again, they fall into
natural classes. We first consider the five two-vertex graphs in which an even
number of bosons are exchanged between the pairs of $\alpha$ vertices
$(v_1,~v_3)$ and $(v_5,~v_7)$. For a given $l$ the sum of the five amplitudes is
\begin{eqnarray}
&&{v_1v_5\Lambda^{2\alpha+2\gamma+1-2l}\int dx\,\Delta^{2l}\over
(\alpha+1-l)!(\alpha+\gamma-l)!(2l)!2^{2\alpha+\gamma+1-2l}}
-{v_1v_7g\Lambda^{2\alpha+2\gamma+1-2l}\int dx\,\Delta^{2l}\over
(\alpha+1-l)!(\alpha+\gamma-1-l)!(2l)!2^{2\alpha+\gamma-2l}}\nonumber\\
&&-{v_3v_5g\Lambda^{2\alpha+2\gamma+1-2l}\int dx\,\Delta^{2l}\over
(\alpha-l)!(\alpha+\gamma-l)!(2l)!2^{2\alpha+\gamma-2l}}
+{v_3v_7g^2\Lambda^{2\alpha+2\gamma+1-2l}\int dx\,\Delta^{2l}\over
(\alpha-l)!(\alpha+\gamma-1-l)!(2l)!2^{2\alpha+\gamma-1-2l}}\nonumber\\
&&-{v_3v_7\Lambda^{2\alpha+2\gamma-1-2l}\int dx\,\Delta^{2l}\mbox{Tr}\,
\bar S(x)S(x)\over(\alpha-l)!(\alpha+\gamma-1-l)!(2l)!2^{2\alpha+\gamma-2l}}.
\label{e4.17}
\end{eqnarray}

There are five more graphs in which an odd number of bosons are exchanged.
The sum of the amplitudes of these graphs is similar to the result in
Eq.~(\ref{e4.17}) and we do not give it here. Furthermore, there are ten
corresponding graphs in which $\alpha$ is replaced by $\beta$ and these are
constructed from the vertices $v_2$, $v_4$, $v_6$, and $v_8$.
When these twenty amplitudes are combined and summed over $l$ no dramatic
cancellation like that in the calculation of the ground-state energy density
occurs. Thus, we are left with an infinite sum over integrals of the
coordinate space propagator $\Delta(x)$.

Next, we consider the contributions of the twenty graphs, analogous to those
above, that are constructed from one $\alpha$ vertex and one $\beta$ vertex.
That is, we construct all possible multi-boson exchange graphs from the vertices
$(v_1,~v_3)$ connected to $(v_6,~v_8)$, and $(v_2,~v_4)$ connected to
$(v_5,~v_7)$. This calculation simplifies dramatically because only one- and
two-particle exchange graphs survive when we apply the derivative operator in
Eq.~(\ref{e2.12}) and set $\alpha=\beta=0$ and $\gamma=1/2$.

Last, we include the four single-vertex graphs constructed from the vertices
$v_5$, $v_6$, $v_7$, and $v_8$.

When we combine all of these calculations we obtain the following result:
\begin{eqnarray}
&&\langle\phi\rangle=i\sqrt{\pi\Lambda/2}\Biggm\{-\delta+\delta^2\Biggm[
1-\ln 2\nonumber\\
&&+g^2\int dx\,\Delta(x)\left({\Delta(x)\over2\Lambda}\left[\ln\left({\Lambda
\over2}\right)+\psi(1)\right]+\left[1+{\Delta(x)\over\Lambda}\right]\ln\left[1+
{\Delta(x)\over\Lambda}\right]\right)\Biggm]+{\rm O}(\delta^3)\Biggm\}.
\label{e4.18}
\end{eqnarray}
Note that any positive integer power of the propagator $\Delta(x)$ in
Eq.~(\ref{e4.prop})	is integrable. However, the value of the propagator at
the origin $\Lambda=\Delta(0)$ is a divergent quantity. Therefore, the function
$\Delta(x)/\Lambda$ vanishes everywhere except at $x=0$, where it is unity.
Hence, the integral involving this ratio in Eq.~(\ref{e4.18}) exists and
vanishes. Thus, our final result for the one-point function, which measures the
parity symmetry breaking in this theory, is
\begin{eqnarray}
\langle\phi\rangle=i\sqrt{\pi\Lambda/2}[-\delta+\delta^2(1-\ln 2)+{\rm O}(
\delta^3)].
\label{e4.19}
\end{eqnarray}

The fact that the theory is supersymmetric reduces the degree of divergence
of this result. At intermediate stages of the calculation, the coefficient of 
$\delta^2$ is proportional to $\Lambda^{1/2}\ln\Lambda$. However, when the boson
and fermion contributions are combined, all terms containing $\ln\Lambda$ cancel
exactly. Thus, the higher-order result is no more divergent than the
leading-order result.

\subsection{Calculation of the fermion-boson mass ratio to second order in
$\delta$}
\label{ss4c}

We do not discuss this calculation in detail here because a similar one is
explained in Ref.~\cite{SUPER2}. The calculation done here is more elaborate
because there are twice as many vertices but the necessary calculations are
routine. Our result is
\begin{eqnarray}
R={M_{\rm fermion}\over M_{\rm boson}}=1+{\rm O}(\delta^3),
\label{e4.20}
\end{eqnarray}
which is consistent with the theory being supersymmetric.

\section {CONCLUSIONS}
\label{s5}

In the Schwinger model of two-dimensional electrodynamics with massless fermions
there is an anomaly. If the one-fermion-loop contribution to the photon
propagator is calculated naively one obtains a product of two factors; the first
factor vanishes in two-dimensional space, and the second factor is a divergent
integral. If one is not careful, one gets a quantity that is formally $0$.
However, because the integral is divergent, one must evaluate this product by
introducing a regulator; dimensional regulation is an acceptable procedure. As
the regulator is removed one obtains a finite, nonvanishing result for the
anomaly, namely, the famous number $e^2/\pi$. In general, one looks for an
anomaly when there is a naive cancellation involving divergent quantities that
must be regulated. The question that is raised in this paper is, do we have an
anomaly in the $\delta$ expansion that breaks supersymmetry? Specifically, is
there an anomaly associated with the cancellation that gives a vanishing
ground-state energy density in Eq.~(\ref{e3.7})?

\subsection{Dimensional regularization of the ground-state energy density}
\label{ss5a}

In the derivation of Eq.~(\ref{e3.7}) we combine two numbers that are divergent
to obtain 0. There are several ways to regulate the integral representing
$\Lambda$. For example, if we use dimensional regulation and evaluate $\Lambda$
in $2-\epsilon$ dimensions, then for small positive $\epsilon$
\begin{eqnarray}
\Lambda &\sim& {1\over 2\pi}\int_0^{\infty}dp\,{p^{1-\epsilon}
\over p^2+g^2} \quad (\epsilon\to 0+)\nonumber\\
&=& {1\over4\pi}g^{-\epsilon}\int_0^1du\,u^{-1+\epsilon/2}
(1-u)^{-\epsilon/2}\nonumber\\
&=& {1\over4\pi}g^{-\epsilon}\Gamma\left({\epsilon\over2}\right)\Gamma
\left(1-{\epsilon\over2}\right)\nonumber\\
&=& {1\over4}g^{-\epsilon}{1\over\sin(\pi\epsilon/2)}\sim{1\over2\pi\epsilon}
g^{-\epsilon}\quad(\epsilon\to0+).
\label{e5.1}
\end{eqnarray}
Furthermore, in $d$-dimensional space, the representation of the Dirac matrices
has rank $2^{d/2}$. Thus, the trace of a unit matrix in $2-\epsilon$ dimensions
is
\begin{eqnarray}
{\rm Tr}{\openone}=2^{(2-\epsilon)/2}\sim 2-\epsilon\ln2\quad (\epsilon\to0+).
\label{e5.2}
\end{eqnarray}
Thus, the coefficient of the second graph amplitude in Eq.~(\ref{e3.6}) should
be multiplied by $1-{1\over2}\epsilon\ln2$.

We now see that the two graphs do not exactly cancel; rather, the difference in
the numerical coefficients is of order $\epsilon$. Combining the two graphs in
(\ref{e3.6}) now gives
\begin{eqnarray}
\epsilon(\ln2)\delta g^2{(2\alpha+1)!\Lambda^{\alpha+1}\over2^{\alpha+2}\alpha!}
&=& {\epsilon\ln2\over2\sqrt{\pi}}\delta g^2 2^\alpha\Gamma\left(\alpha
+{3\over2}\right)\Lambda^{\alpha+1}\nonumber\\
&\sim& {\ln2\over4\pi^{3/2}}\delta g^2\Gamma\left(\alpha+{3\over2}\right)
(2\Lambda)^{\alpha}\quad(\epsilon\to0+).
\label{e5.3}
\end{eqnarray}
If we now differentiate with respect to $\alpha$ and set $\alpha=0$, we obtain
for the ground-state energy density
\begin{eqnarray}
E_{\rm ground~state} &=& {\ln2\over8\pi}\delta g^2\left[\psi\left({3\over2}
\right)+\ln(2\Lambda)\right]+{\rm O}(\delta^2)\nonumber\\
&\sim& {\ln2\over8\pi}\delta g^2\left[\psi\left({3\over2}\right)
-\ln(\pi\epsilon)\right]+{\rm O}(\delta^2)\quad (\epsilon\to0+).
\label{e5.4}
\end{eqnarray}
Because this is a positive number, it suggests that supersymmetry may be broken.

Of course, dimensional regulation violates supersymmetry. Thus, it is not clear
whether the nonzero result in Eq.~(\ref{e5.4}) is correct or is merely an
artifact of the regularization scheme being used.

\subsection{Proper-time regularization}
\label{ss5b}

If a supersymmetric regulation exists then, of course, there will be no anomaly.
Conversely, if we could establish rigorously that there does not exist any
supersymmetric regulation of the delta expansion, then there really would be a
breaking of supersymmetry. We do not know for certain whether a supersymmetric
regulation of the $\delta$ expansion exists. However, we believe that we have
found a relatively simple way to regulate the $\delta$ expansion that is
consistent with supersymmetry and thus we believe that there is no anomalous
structure and that the ground-state energy is truly identically zero. This
suggests that while it is relatively easy to break parity symmetry,
supersymmetry is extremely rigid and is very difficult to break.

Our regulation scheme, which we believe respects supersymmetry invariance, is a
variant of the proper-time method due to J.~Schwinger \cite{SCHW}. (It is well
known that the proper-time method correctly yields the anomaly in the Schwinger
model.) Here, we will use this method to define, in an invariant way, the
divergent integral $\Lambda$. To make contact with the dimensionally regulated
result above, let us work in $d$ dimensions:
\begin{eqnarray}
\Lambda=\int{d^dp\over(2\pi)^d}{1\over p^2+g^2}=\int{d^dp\over(2\pi)^2}
\int_{s_0}^\infty ds\,e^{-s(p^2+g^2)}.
\label{e5.5}
\end{eqnarray}
Here, $s_0$ is a proper-time regulator to be taken to 0 at the end of the
calculation. We now interchange the order of integration in (\ref{e5.5}), and
express the momentum integral as the product of $d$ one-dimensional integrals:
\begin{eqnarray}
\int_{-\infty}^\infty {dp\over2\pi}e^{-sp^2}={1\over2\sqrt{\pi s}}.
\label{e5.6}
\end{eqnarray}
Then, the integral representing $\Lambda$ is immediately expressed in terms of a
single regulated integral:
\begin{eqnarray}
\Lambda=\int_{s_0}^\infty ds\,e^{-sg^2}\left(1\over2\sqrt{\pi s}\right)^d
={g^{d-2}\over 2^d\pi^{d/2}}\int_{g^2s_0}^\infty dx\,x^{-d/2}e^{-x}.
\label{e5.7}
\end{eqnarray}
If we set $d=2-\epsilon$ as in the previous subsection, the integral converges
when $s_0=0$, and we obtain the same result as in (\ref{e5.1}),
\begin{eqnarray}
\Lambda\sim{1\over4\pi}\Gamma\left(\epsilon\over2\right).
\label{e5.8}
\end{eqnarray}
However, if we want to preserve supersymmetry, we must remain in two dimensions,
in which case the integral depends logarithmically upon $s_0$:
\begin{eqnarray}
d=2:\quad\Lambda={1\over4\pi}\int_{gs_0}^\infty{dx\over x}e^{-x}\sim
-{1\over 4\pi}[\gamma+\ln(gs_0)],
\label{e5.9}
\end{eqnarray}
where $\gamma$ is Euler's constant.

The obvious advantage of this regulation scheme is that it treats bosons and
fermions on an equal footing; the renormalization of the boson and fermion
masses is identical. With this regulation scheme, the ground-state energy
density is zero, as expected. Thus, we believe that there is no anomaly in the
$\delta$ expansion.

\section* {ACKNOWLEDGEMENT}
\label{s6}

We thank A.~Das, L.~Gamberg, and M.~Grisaru for illuminating discussions, and we
are grateful to the U.S.~Department of Energy for financial support.

\begin{figure}
\caption{The two graphs contributing to the ground-state energy to first order
in $\delta$. These graphs are constructed from vertices having an even number of
boson lines. The symmetry numbers are shown beside each graph.}
\label{fig1}
\end{figure}
\begin{figure}
\caption{The two graphs contributing to the one-point Green's function to first
order in $\delta$. These graphs are constructed from vertices having an odd
number of boson lines. The symmetry numbers are shown beside each graph.}
\label{fig2}
\end{figure}
\begin{figure}
\caption{One-particle-irreducible graphs contributing to the mass
renormalization of the fermion and the boson to order $\delta$. There is one
graph for the fermion mass shift and two graphs for the boson mass shift. All
graphs are constructed from vertices having an even number of boson lines. The
symmetry numbers for the graphs are shown beside each graph.}
\label{fig3}
\end{figure}
\begin{figure}
\caption{The thirty graphs contributing to the ground-state energy to second
order in $\delta$. These graphs are constructed from the eight vertices in
Eq.~(4.3). We have organized the graphs into sets for which the amplitudes
combine in a natural way. These sets are labeled (a)--(g).}
\label{fig4}
\end{figure}
\begin{figure}
\caption{Five of the graphs contributing to $\langle\phi\rangle$ to second order
in $\delta$. These graphs are constructed from the eight vertices in Eq.~(4.3).}
\label{fig5}
\end{figure}
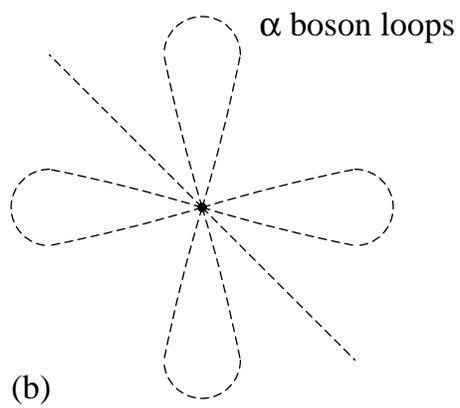
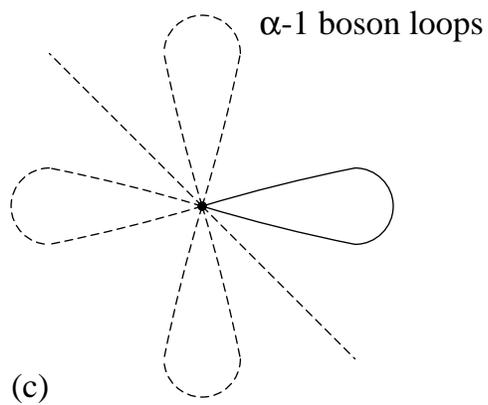
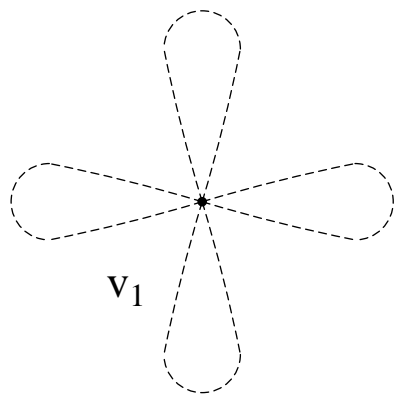
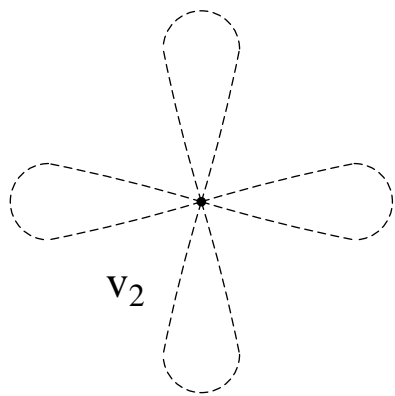
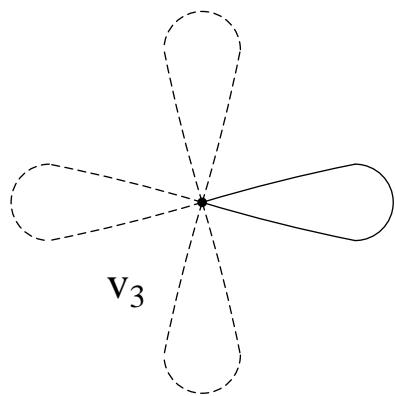
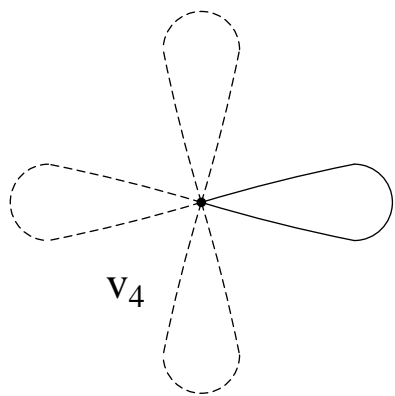
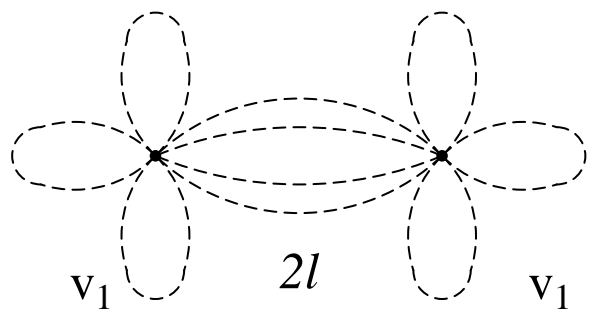
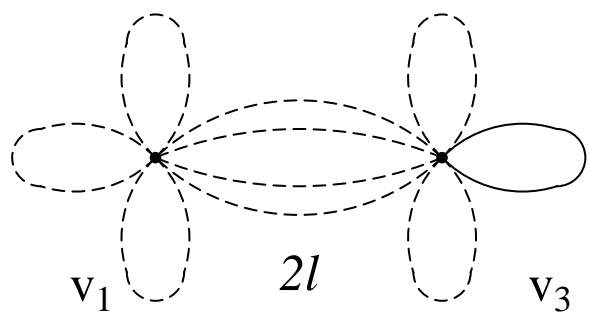
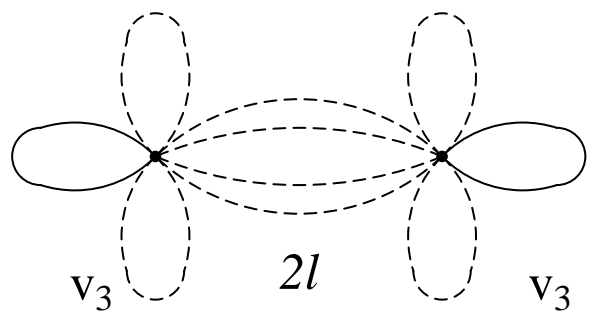
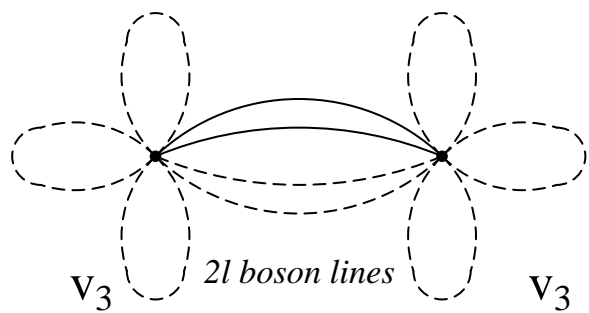
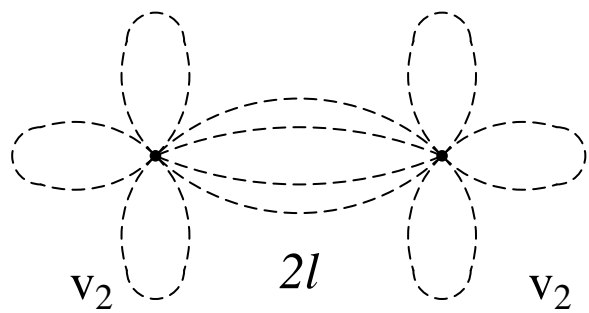
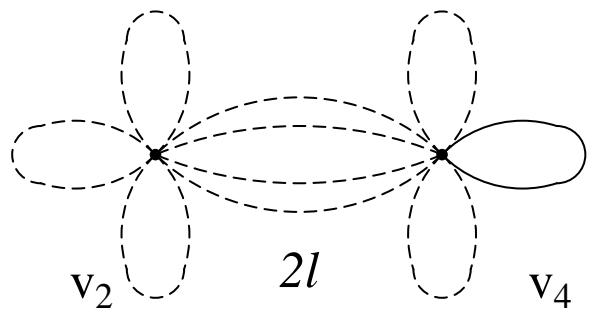
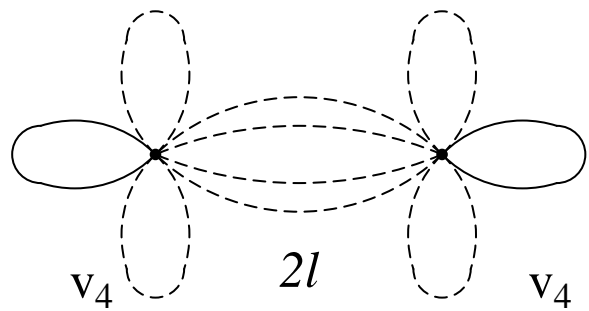
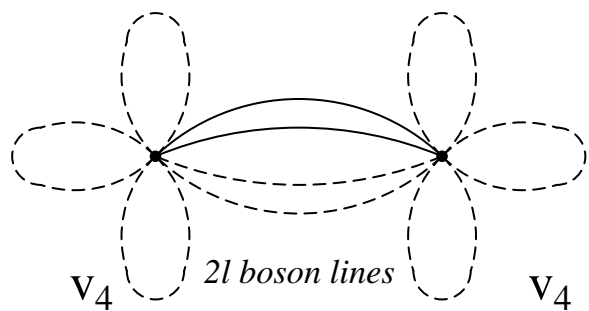
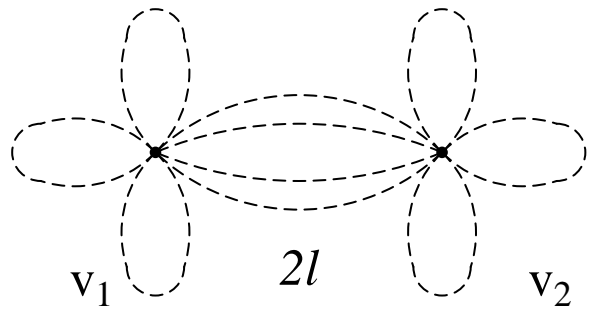
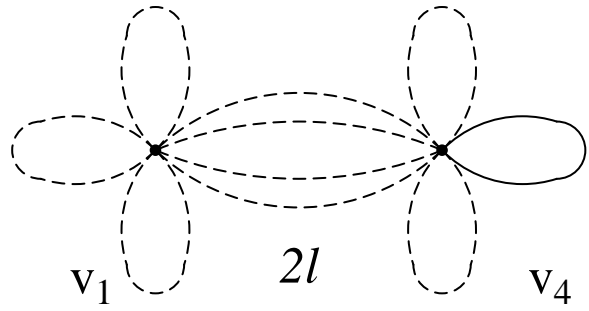
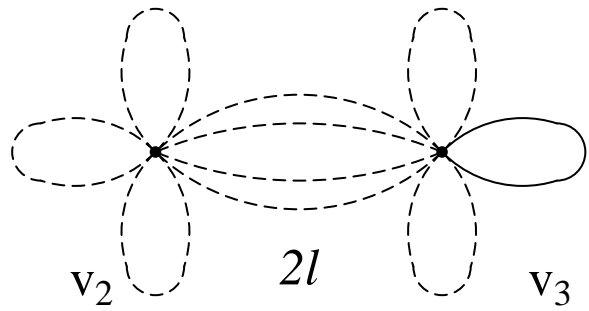
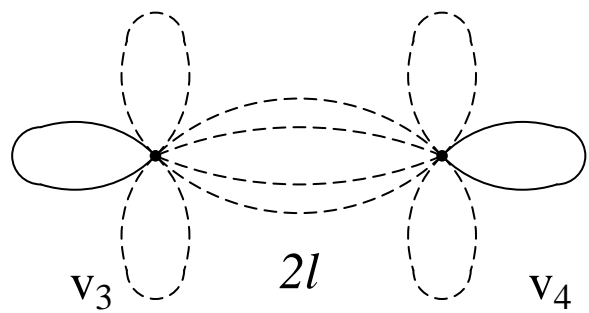
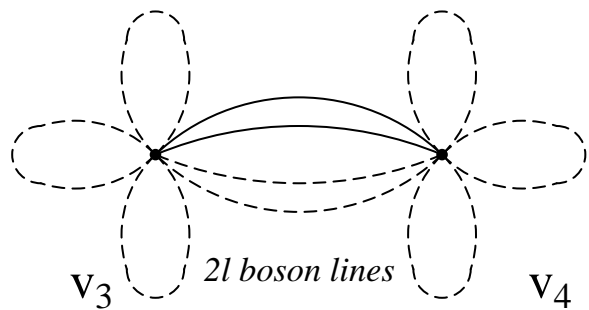
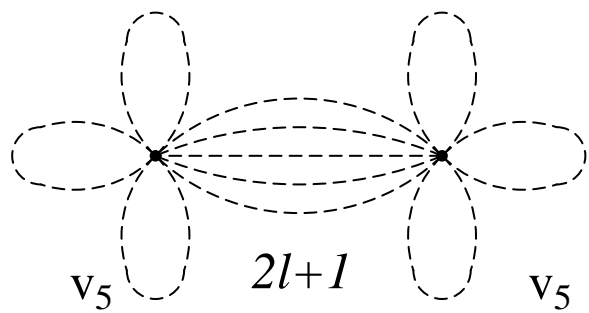
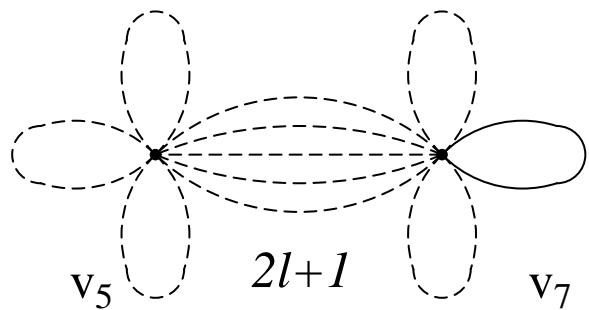
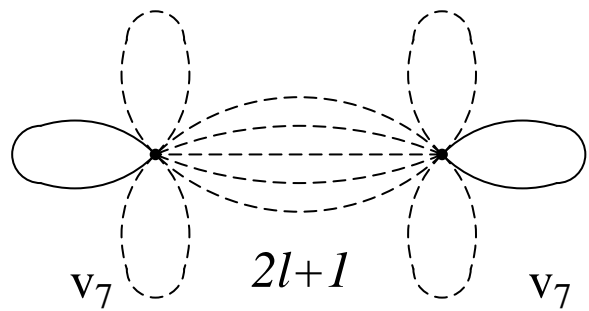
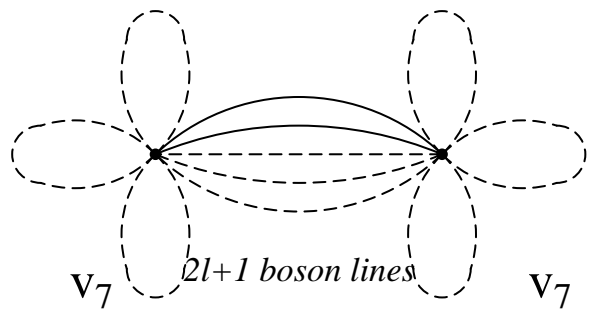
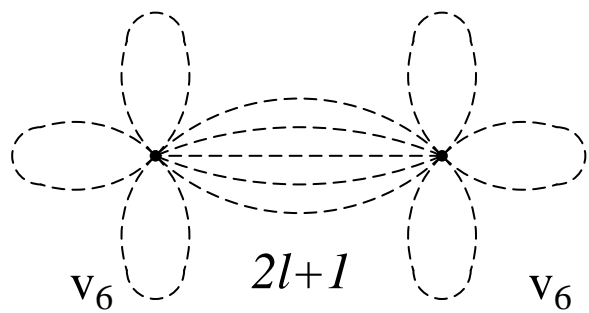
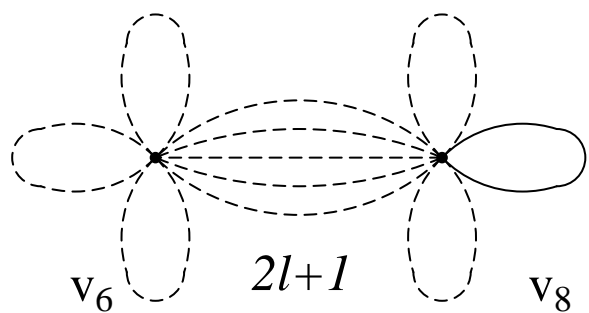
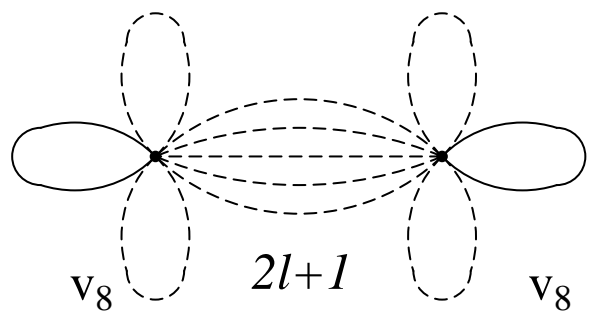
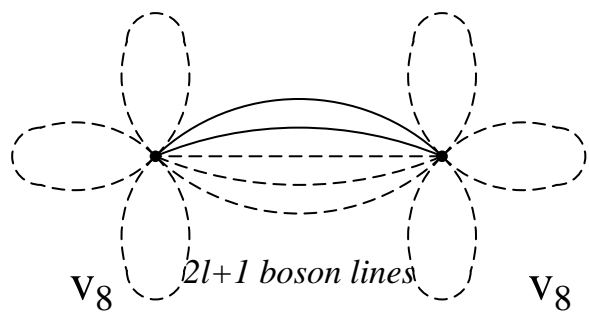
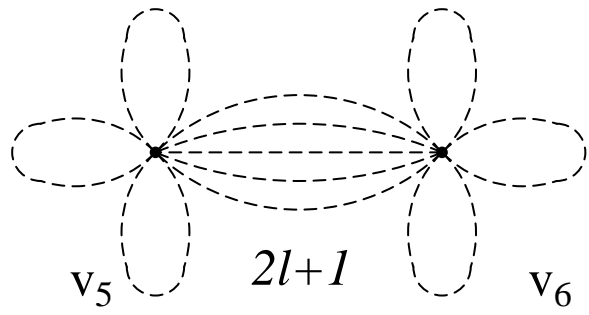
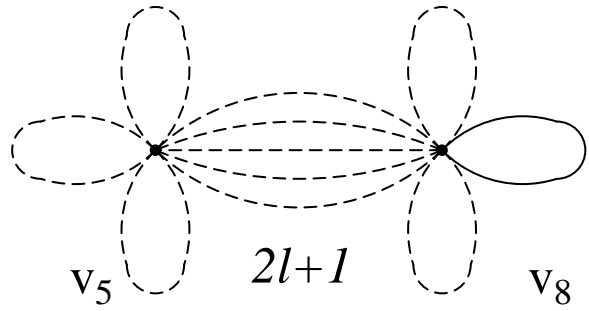
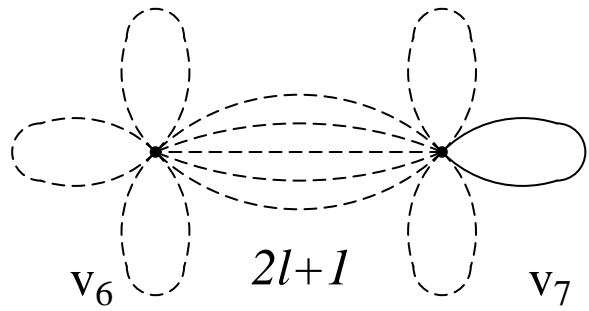
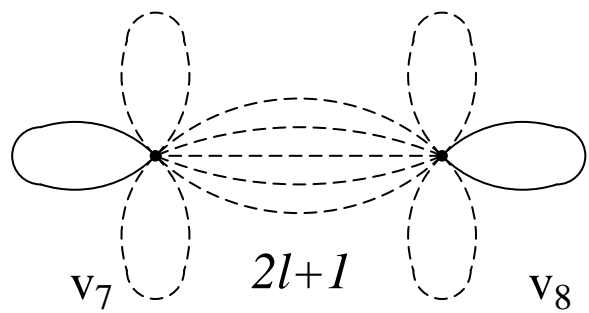
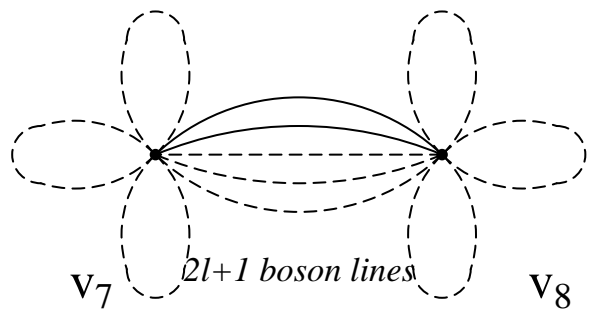
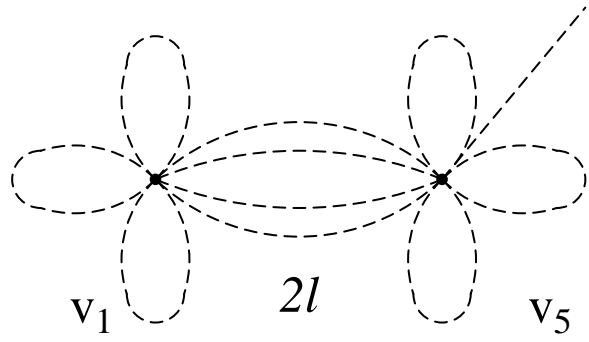
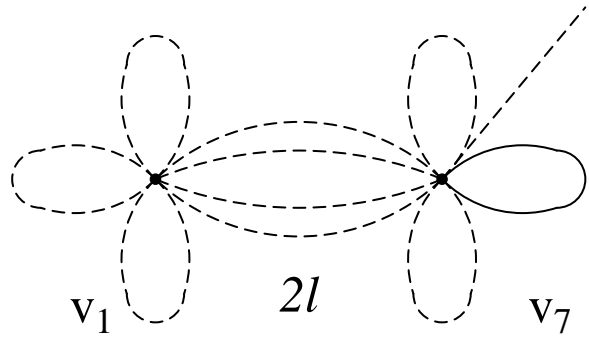
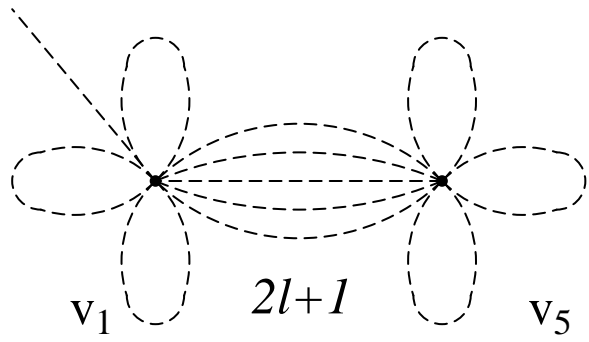
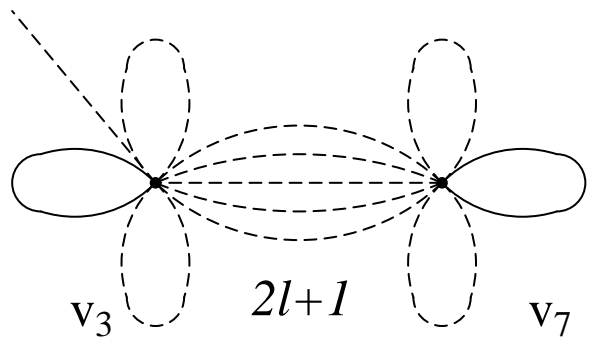
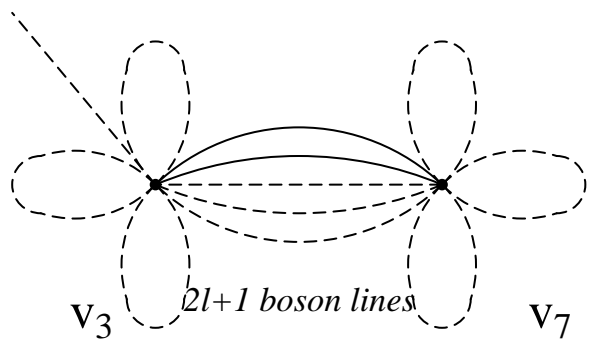
\end{document}